\documentclass[a4paper]{article}
\usepackage{amsmath}
\usepackage{amsthm}
\usepackage{comment}
\usepackage{graphicx} 
\usepackage[utf8]{inputenc}
\usepackage[capitalise]{cleveref}
\usepackage{xcolor}
\usepackage[subrefformat=parens]{subcaption}
\usepackage{url}
\usepackage{multirow}

\usepackage{nccmath} 

\usepackage {vtable}
\usepackage{bm}
\usepackage{natbib} 
\usepackage[margin=25truemm]{geometry}
\everymath{\displaystyle}

\usepackage{verbatim}

\title{Understanding changes in traffic demand during the Tokyo 2020 Olympic and Paralympic Games }  
\author{Takao Dantsuji\thanks{Corresponding author, Monash Institute of Transport Studies, Monash University, 23 College Walk, Clayton, Victoria 3800, Australia,  Takao.Dantsuji@monash.edu} \thanks{Institute of Science and Engineering, Kanazawa University, Kakuma-machi, Kanazawa, Ishikawa 920-1192, Japan}  \and Masaki Nakagawa \thanks{Graduate School of Natural Science and Technology, Kanazawa University, Kakuma-machi, Kanazawa, Ishikawa 920-1192, Japan, masaki.nakagawa.1191@gmail.com}} 
\date{}

\begin{document}
\maketitle

\begin{abstract}
    This paper evaluated the effects of the Tokyo 2020 Olympic and Paralympic Games on traffic demand on the Metropolitan expressway. We constructed panel data for both passenger and freight vehicles' demand using longitudinal disaggregated trip records from the Metropolitan expressway. Subsequently, we established a demand function and used a difference-in-differences method to individually estimate the impacts of toll surcharges and other Olympics-related factors by leveraging the fact that the toll surcharges were not applied to freight vehicles.

    The results indicate that toll surcharges resulted in a decrease of $25.0\%$ for weekdays and $36.8 \%$ for weekends/holidays in passenger vehicle demand on the Metropolitan expressway. The estimated toll elasticities are $0.345$ for weekdays and $0.615$ for weekends/holidays, respectively. Notably, analysis of the Olympics-related factor demonstrated that travel demand management (TDM)  strategies effectively curbed demand on weekends/holidays with a reduction of $2.9 \%$ in traffic demand. However, on weekdays, induced demand surpassed the reduction of demand by  other TDM strategies than tolling, resulting in a $4.6 \%$ increase in traffic demand.

    Additionally, We developed a zone-based demand function and investigate the spatial heterogeneity in toll elasticity. Our findings revealed small heterogeneity for weekdays ($0.283$ to $0.509$) and large heterogeneity for weekends/holidays ($0.484$ to $0.935$).

    {\flushleft{{\bf Keywords:} Tokyo 2020 Olympic and Paralympic Games; demand; travel demand management; toll surcharges; difference-in-differences.}}
\end{abstract}

\section{Introduction}
\subsection{Background}
Travel demand management (TDM) is crucial for effectively operating the transportation system of host cities during mega events like the Olympic Games and the FIFA World Cup.
Various stakeholders including media personnel, staff, and athletes participate in these events and  their travel needs are prioritized to make the event successful. Beyond regular commuting
and freight transportation, these additional trips stress transportation systems.  The literature underscore the importance of TDM strategies during mega events \cite[e.g.,][]{currie2012synthesis}.

The summer Olympic and Paralympic Games (hereafter refereed to as the Olympic Games) represent one of the largest events worldwide,  where various TDM strategies are implemented.
A review by \cite{currie2011assessing} on the TDM strategies of the Olympic Games between 1980 (Moscow) and 2016 (Rio de Janeiro) revealed system-wide base load demand reductions of up to $30 \%$  and localized impacts up to $40 \%$. While the study showed the effectiveness of TDM strategies during the Olympic Games, the specific effects of each strategy remain unclear. It is useful for policymakers to understand the effects of the TDM strategies for effective implementation.

The Tokyo 2020 Olympic Games were held without spectators after being postponed for one year owing to the COVID-19 pandemic. With a surge in COVID-19 cases preceding  the Games, a state of emergency (SOE) was declared on July 12, 2021, promoting calls for reduced social interactions to mitigate the spread of COVID-19. Although the Games were held entirely under the SOE, a range of TDM strategies were implemented during the Tokyo Olympic Games to ensure the safe and efficient transportation of the Olympic families via the Metropolitan expressway.

\begin{figure}[t!]
  \centering
  \includegraphics[width=\textwidth]{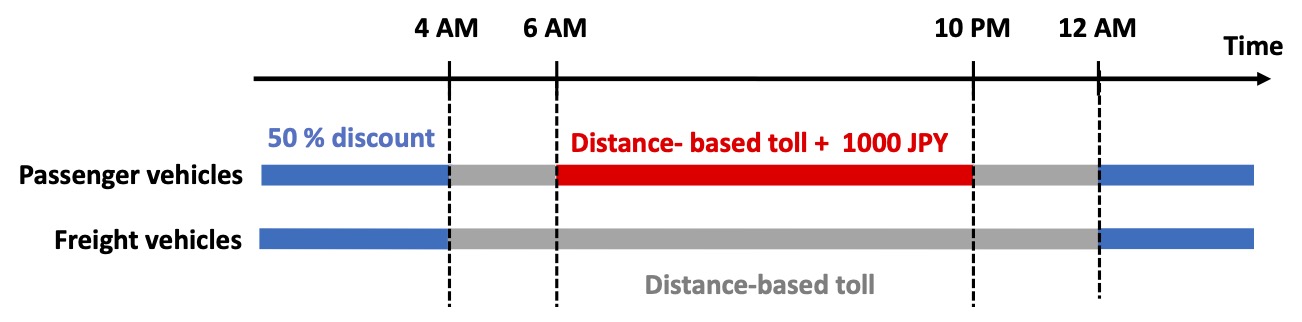}
  \caption{Toll structure of the Metropolitan expressway during the Tokyo Olympic Games}\label{fig:price_figure}
\end{figure}

An important TDM strategy implemented during the Games involved toll surcharges during daytime hours and discounts at night (Fig.~\ref{fig:price_figure}). Passenger cars incurred an additional charge of $1,000$ JPY from 6:00 AM to 10:00 PM, while toll fees for all vehicles were reduced by $50 \%$ from 12:00 AM to 4:00 AM, aiming to shift trips from peak to off-peak times. A notable decrease of approximately $20 \%$ in traffic volume on the Metropolitan expressway during the Games compared to the same period in 2019 was reported \citep{report2022}. However, other factors such as additional TDM strategies and the influence of COVID-19 also contributed to this reduction. Given the potential legacy for time-dependent pricing, a more accurate assessment of its effects on traffic demand is warranted.

This paper evaluated  the impacts of toll surcharges and other Olympics-related factors  on traffic demand separately (hereafter referred to as toll and Olympics effects, respectively). To achieve this objective, we constructed panel data for both passenger and freight vehicles' demand using disaggregated trip records from the Metropolitan expressway. Subsequently, we established a demand function and applied a difference-in-differences (DID) method to estimate their respective effects by leveraging the fact that the toll surcharges were not levied for freight vehicles. Additionally, we developed a zone-based model to investigate the spatial heterogeneity in these effects. 

\vspace{12pt}
The remaining of this paper is organized as follows. Section 2 reviews the Tokyo 2020 Olympic and Paralympic Games and their associated TDM strategies. Section 3 provides the literature review.   Section 4 describes the data used in this paper. The methodology is detailed in Section 5. Section 6 presents the results, concluding this paper in Section 7. 

\section{The Tokyo 2020 Olympic and Paralympic Games}

The Tokyo 2020 Olympic and Paralympic Games were held without spectators from July 23,  2021 to August 8, 2021 and from August 24, 2021 to September 5, 2021, respectively. Prioritizing sustainable development, the Tokyo 2020 Olympic Games maximized the use of existing venues, with only eight new venues constructed out of 43. Most venues were located in the center of Tokyo, as depicted in Fig. \ref{fig:o_map_eng}. Because Tokyo is one of the world's most congested megacities \citep{tomtom}, numerous TDM strategies were implemented to ensure the safe and smooth transportation of Olympic participants while accommodating regular activities such as business and freight trips. This section reviews the strategies relevant to this study based on \cite{report2022}.

\subsection{Toll strategies}

\begin{figure}[t!]
  \centering
  \includegraphics[width=0.8\textwidth]{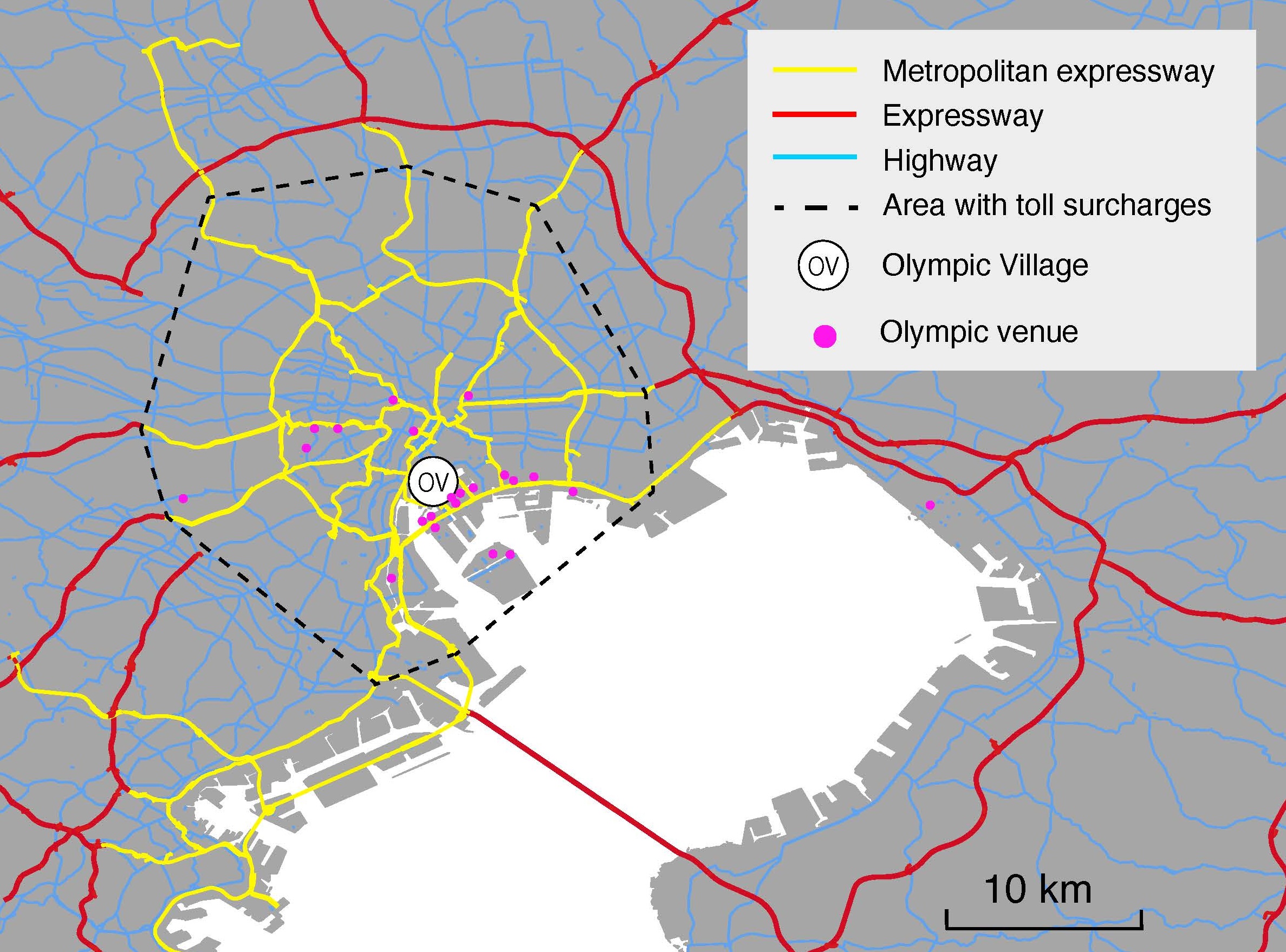}
  \caption{Expressway and highway networks in the Tokyo Metropolitan Area }\label{fig:o_map_eng}    
\end{figure}

{One important strategy involved the toll surcharges on a specific part of  Metropolitan expressway as shown in Fig.~\ref{fig:o_map_eng}. An additional 1,000 JPY was imposed on passenger cars from 6:00 AM to 10:00 PM, irrespective of trip length, while toll fees for all vehicles were reduced by 50 \% from 12:00 AM to 4:00 AM\footnote{It is important to note that the discount applied only to vehicles equipped with the electric toll collection (ETC) system and not to vehicles whose drivers paid by cash} to encourage a shift in trip timing  from peak to off-peak hours.  The Metropolitan expressway's minimum and maximum usual distance-based tolls  for passenger  vehicles were 300 and 1,320 JPY, respectively. These toll strategies were implemented four days before the Olympic Games and ended one day after the Olympic Games. The period of toll strategies for the Paralympic Games coincided with that of the Paralympic Games. We consider the vehicle types for which the toll surcharges were levied as ``passenger vehicles" and the other vehicle types are ``freight vehicles" in this study.  }

{According to the 6th Tokyo Metropolitan Person Trip Survey \citep{PT}, 65.5 \% of the Tokyo residents use railway for commuting purposes, while only 8.0 \% of them commute by car. Railways are dominant travel mode for commuting in Tokyo. For business purposes, 42.9 \% and 37.7 \% of residents travel by railways and car, respectively. For personal activity purposes, 16.4 \% travel by car. While we are unsure how many passenger car drivers use the Metropolitan expressway, we deduce from this survey that a large portion of passenger car trips are made for business purposes on weekdays. }

\subsection{Additional strategies implemented on the Metropolitan expressway}
Similar to  recent Olympic Games in Rio de Janeiro, London, Beijing and Athens \cite[][]{currie2011assessing}, an Olympic route network was established by designating lanes exclusively for Olympic families on the expressways and highways near the venues to alleviate traffic congestion. 

Toll plazas, junction lane restrictions, and entry bans were enforced based on expressway traffic conditions. Lane restrictions were imposed at least once at 13 toll plazas, with an average implementation of 18.2 days out of the 22 days of the Olympic Games and throughout the Paralympic Games. Junction lane restrictions were enforced at 
10 junctions for an average of 3.4 days during the Olympic Games and at 6 junctions for an average of 2.3 days during the Paralympic Games. Additionally, toll plazas were closed at 52 locations for an average of 11.5 days during the Olympic Games and at 36 locations for an average of 8.1 days during the Paralympic Games. 

\subsection{Promotion and marketing campaigns}
The Tokyo Metropolitan Government (TMG) incentivized companies to invest in telecommuting equipment or software to encourage business telecommuting and alleviating travel demand during rush hours. Moreover, 70 \% of TMG staff engaged in telecommuting during the Games. Additionally, over 90 \% of the companies and organizations participating in the ``2020 TDM promotion project"\footnote{This project initiated by the TMG, Cabinet Secretariat and Bureau of Tokyo 2020 Olympic and Paralympic Games Preparation aimed to secure cooperation from companies and organizations regarding TDM, with more than 50,000 entities participating in the initiative \cite{report2022}. } introduced  telecommuting, staggered work hours, or scheduled vacations during the Games. Furthermore, logistic companies were encouraged to adopt efficient delivery approaches such as joint delivery and timing shifts through seminars, consultations and workshops organized by the TMG.

\section{Literature review}
\subsection{The impacts of TDM strategies during the Olympic Games}
Existing literature provides insights into  changes in travel behaviors and traffic demand resulting from TDM strategies during the Olympic Games. A previous study \citep{currie2011assessing} conducted a comprehensive review of TDM strategies implemented from 1980 (Moscow) to 2016 (Rio de Janeiro), highlighting their significance. The study found that base load demand could be reduced by up to $30 \%$, and time-specific effects of TDM strategies could decrease demand by as much as $40 \%$. Another study \citep{parkes2016understanding} investigated commuters' travel behavior changes during the London 2012 Olympic Games, showing significant alterations, including trip  reduction and re-timing.  \cite{currie2014travel} reviewed TDM strategies during the same event, noting the ``Big scare" effect, where pre-Games warnings, media attention, and unplanned events influenced travel behaviors in London. \cite{mingjun2008comparison} demonstrated decreased traffic volumes on expressway, arterial roads, secondary roads, and branch roads by 19 \%, 22 \%, 39 \%, and 25 \%, respectively during the Beijing 2008 Olympic Games.  While these studies suggest significant reductions in traffic demand owing to TDM strategies, induced demand is possible. \cite{li2016effects} reported the effects of the ``odd and even" car number plate restriction policy during the Beijing 2008 Olympic Games, indicating that  despite restricting over $50$ \% of vehicles, traffic volume only decreased by $20 \%$ -- $40 \%$. The authors suggested that the policy might have increased trip lengths or induced demand on the main roads.

\subsection{DID applications in the transportation field}
{DID is a statistical method commonly used to estimate causal effects by comparing changes in outcomes over time between a treatment group and a control group. DID methods, frequently employed for analyzing the causal effects of transportation policies through quasi-experiments, have been utilized in various studies  \cite[e.g.,][]{foreman2016crossing, liang2023short}. For instance, DID methods have been used to evaluate the effects of TDM strategies such as pricing \cite[e.g.,][]{graham2020quantifying, zheng2023impacts, liang2023short, rodseth2024panel} and vehicle driving restriction \cite[e.g.,][]{liang2023panacea}, and the effects of changes in the built environment,  such as new metro lines \cite[e.g.,][]{deng2022impact}, new railways \cite[e.g.,][]{wang2021heterogeneous, wang2024high}, and  new stations \cite[e.g.,][]{rojas2024train}.  \cite{kuo2021impact}  investigate the causal impacts of a large-scale event (Pokemon Go Safari Zone event) on bikeshare systems and show that the event increased bikeshare trips by 27\%.}

\subsection{Research gaps} 
The existing literature leaves several important questions. First, 
many studies compare traffic metrics such as flow and demand before and during the Games to estimate the overall impacts of TDM strategies as a net effect. Unfortunately, the individual effects of each TDM strategy were not thoroughly explored. 
Understanding how travelers respond to each TDM strategy is crucial for effective TDM implementation.   Second, these analyses may not capture the causal effect of TDM strategies and could be biased by various factors such as weather and the day of the week. Establishing a more accurate understanding of the impacts of TDM strategies requires investigating the individual causal effects of each strategy. Third, although \cite{li2016effects} suggests the possibility of induced demand, the existence of induced demand was not fully verified in the existing literature. Further investigation is required to determine whether TDM strategies could potentially increase overall traffic demand. {Fourth,  although existing studies demonstrate that mega-events substantially change transportation system usage using a DID approach \citep{kuo2021impact}, the causal effects of such events on traffic demand, which is closely related to transportation planning \citep{dantsuji2022novel}, have not been sufficiently explored in the literature. To the best of our knowledge, this is the first study to investigate the causal impacts of a mega event, including TDM strategies, on traffic demand using a DID method.}

Addressing these gaps in the literature would contribute to a more comprehensive understanding of the effectiveness and potential side effects of TDM strategies during mega events.

\begin{figure}[!t]
      \centering
  \includegraphics[width=\textwidth]{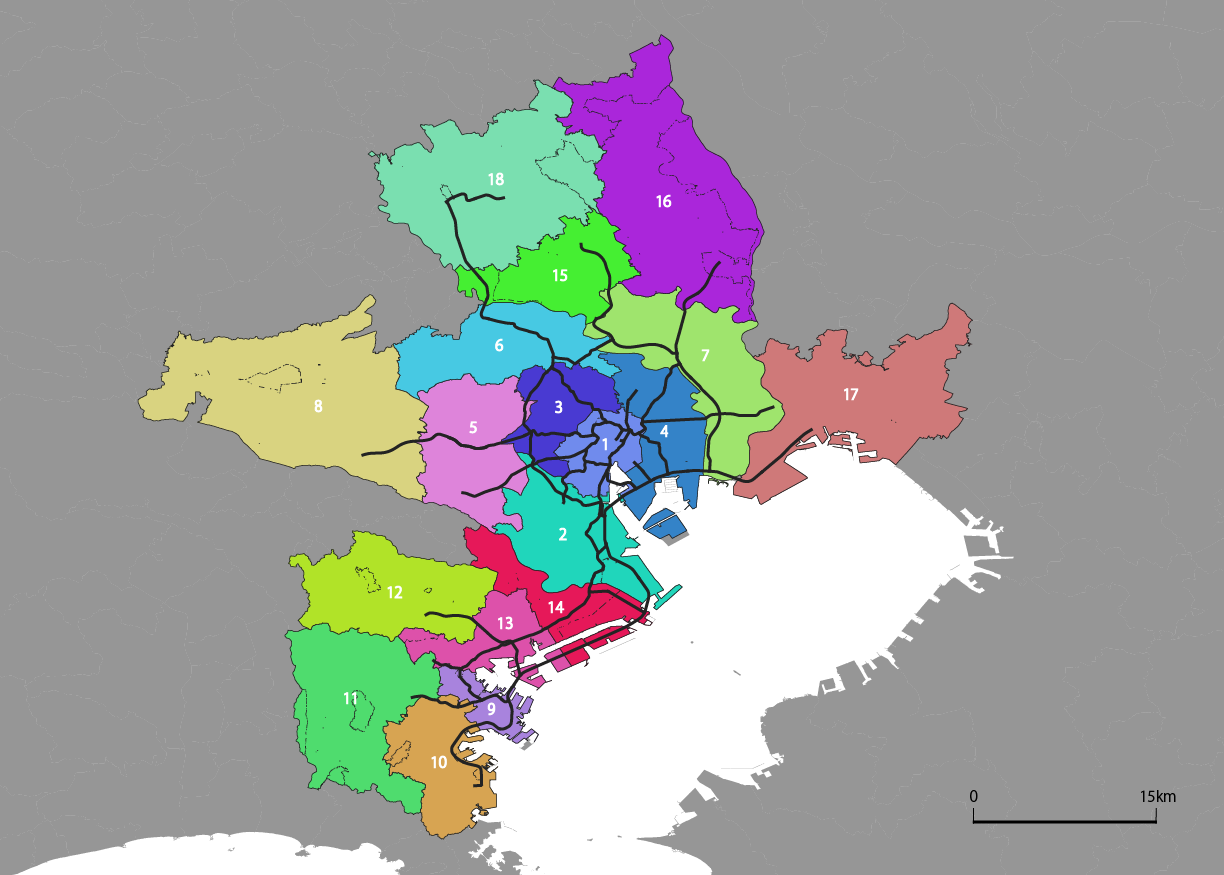}
  \caption{Tokyo Metropolitan expressway network and zone classification}\label{fig:zone_classification}
\end{figure}

\section{Data}
 {We utilize log data from the Electronic Toll Collection (ETC) system, from July 1st to September 30th in 2018, 2019, and 2021 to construct panel data on traffic demand for the Metropolitan expressway (referred to as ETC data). The  ETC system in Japan is an automated toll collection system used on highways and  expressways. Drivers of vehicles equipped with an onboard unit (OBU) that reads the ETC card can pass through toll plazas without stopping via the communication between the OBU and the ETC system. The ETC is widely used across Japan, with most tolled roads, including highways and expressways, supporting the system. With 98.1 \% of vehicles on the Metropolitan expressway using the ETC system as of September
2023, the ETC data provides a rich and sufficient representation of the OD demand of the Metropolitan
expressway. Through this system, information for each vehicle, including vehicle type, origin and destination toll plazas, departure and arrival times, and recorded toll prices, is collected.} 

The daily average of recorded trips is 858,043.2 [veh] as shown in Table \ref{table:descritive_stat}. There are 22,336 OD pairs with existing trip records, and toll surcharges were applied during the Games for  19,827 OD pairs. Therefore, we focus on the trip records of these 19,827 OD pairs to estimate the toll and Olympics effects.

To construct the panel data,  we first aggregate the trip records during the toll surcharge period (i.e., between 6 AM and 10 PM) by year, date, and vehicle types ($D_{y,d,v}$, where $y \in \{ 2018, 2019, 2021 \} $ is the year, $d$ is the date, and $v$ is the vehicle type -- passenger or freight). Additionally we estimate the average toll price $P_{y,d,v}$ corresponding to the demand $D_{y,d,v}$.  Subsequently, we divide the trip records into zonal generated demand ($D^{z}_{y,d,v}$ where $z$ is the zone where demand is generated) to explore spatial heterogeneity. The zone-based average price $P^{z}_{y,d,v}$ is also estimated accordingly. We divide the area into zones based on the zone classification (Fig. \ref{fig:zone_classification}), commonly utilized in travel surveys such as the Tokyo Metropolitan person trip survey,  which considers geographical and historical similarities.  Descriptive statistics and the number of toll plazas where vehicles can enter or exit in each zone are presented in Table \ref{table:descritive_stat}.

\begin{table}[p]
  \caption{Descriptive statistics for zonal demand derived from ETC data and the entry points within zones. Values in parentheses represent the demand by passenger vehicles}
  \label{table:descritive_stat}
  \centering
  \begin{tabular}{lllllll}
    zone & mean & median & max & min & standard deviation & entrance points  \\
    \hline \hline 
 zone 1 &  96480.5 & 101282.5 & 134212 & 46056 & 22827.4 & 33 \\ 
  &  (54456.8) & (54203) & (74841) & (27389) & (11220.2) \\ \hline
 zone 2 &  61970.7 & 67182 & 80190 & 31436 & 12801.2 & 17\\ 
  &  (26897.1) & (27240) & (34481) & (16385) & (3896.8) \\ \hline
 zone 3 &  49068.9 & 51705 & 62254 & 27578 & 8877.3 & 13\\ 
  &  (29284.7) & (29627.5) & (36338) & (18228) & (4679.7) \\ \hline
 zone 4 &  61825.4 & 67404.5 & 82622 & 29726 & 13420.1 & 16\\ 
  &  (28922.6) & (29515) & (38347) & (15885) & (4932.4) \\ \hline
 zone 5 &  70962.9 & 71665.5 & 89489 & 43162 & 10633.3 &8 \\ 
  &  (41456.0) & (42888) & (59624) & (28975) & (7659) \\ \hline
 zone 6 &  32472.3 & 32172 & 42717 & 16424 & 6651.2 & 10 \\ 
  &  (17667.1) & (18465.5) & (25119) & (10889) & (3578.5) \\ \hline
 zone 7 &  80567.0 & 83826.5 & 103102 & 39082 & 15011.4 & 16\\ 
  &  (43215.0) & (44297.5) & (60566) & (26256) & (7144.3) \\ \hline
 zone 8 &  30007.5 & 31017 & 35813 & 18652 & 3301.5 & 1\\ 
  &  (17017.1) & (16318) & (27349) & (11874) & (3262.5) \\ \hline
 zone 9 &  33479.5 & 36954 & 42474 & 16558 & 6998.7 & 13\\ 
  &  (19774.7) & (20243.5) & (23946) & (12277) & (2321.4) \\ \hline
 zone 10 &  28635.0 & 28165 & 37314 & 15614 & 5120.7 & 10\\ 
  &  (17383.8) & (18062.5) & (25890) & (11532) & (2906.1) \\ \hline
 zone 11 &  40238.9 & 42858 & 51195 & 22201 & 7088.7 & 3\\ 
  &  (21933.4) & (21855.5) & (29643) & (15851) & (2211.2) \\ \hline
 zone 12 &  33716.7 & 36185 & 42838 & 16344 & 6799.1 & 5\\ 
  &  (18050.2) & (18125.5) & (23915) & (11918) & (2102.2) \\ \hline
 zone 13 &  24982.1 & 27247 & 32953 & 11113 & 5969.7 & 9\\ 
  &  (11298.3) & (11402.5) & (14551) & (7115) & (1536.8) \\ \hline
 zone 14 &  53847.5 & 56752 & 68881 & 24577 & 9773.5 & 12\\ 
  &  (25335.8) & (25253) & (39179) & (14930) & (4367.2) \\ \hline
 zone 15 &  54738.0 & 57538.5 & 67035 & 30429 & 8436.5 &9 \\ 
  &  (27743.2) & (27313.5) & (42561) & (20128) & (4449.7) \\ \hline
 zone 16 &  39075.3 & 40632.5 & 46979 & 20682 & 5612.9& 3 \\ 
  &  (18966.6) & (18183.5) & (31472) & (12701) & (3842.7) \\ \hline
 zone 17 &  48070.4 & 50405.5 & 60483 & 23881 & 7952.7& 5 \\ 
  &  (24133.6) & (23537) & (35550) & (15151) & (4005.8) \\ \hline
 zone 18 &  21826.9 & 22779.5 & 27299 & 11882 & 3950.8 & 7\\ 
  &  (12960.3) & (13106.5) & (17400) & (8417) & (1875.2) \\ \hline
 Total &  858043.2 & 898607 & 1092361 & 444948 & 154277.8& 180  \\ 
 &  (455059.3) & (464804) & (614093) & (289422) & (64439.5) \\ \hline
  \hline 
  \end{tabular} 
\end{table}

\section{Methodology}

Using the constructed panel data, we employ the DID method to examine the toll and Olympics effects on demand separately. Leveraging the fact that the 1000 JPY toll surcharges were not applied to freight vehicles, passenger vehicles are treated as the treatment group affected by the intervention (the toll surcharges in this paper), while freight vehicles serve as the control group. We assume that  passenger and freight vehicles are subject to the equivalent Olympics effect.  First, we formulate a demand function for the Metropolitan expressway (referred to as the general model hereafter). We then extend the general model to a zone-based model to investigate spatial heterogeneity in these effects. Subsequently, we develop the first-difference model to maintain the assumptions required by the DID method.

\subsection{General model: demand function for the Metropolitan expressway} 
Considering the daily demand dataset of vehicle type $v=\{p,f\}$, where $p$ represents passenger vehicles and $f$ represents freight vehicles, on day $d$ in year $y$, we define the demand function as follows. 

\begin{align}\label{eq:demand_function}
\ln D_{y,d,v} = \alpha_y + \beta_y \delta_{v} + \beta_{p} \delta_{v} \ln P_{y, d, v} + \beta_{ph} \delta_{v} \delta_{h}  \ln P_{y, d, v} + \beta_{o} \delta_{o} +  \beta_{oh} \delta_{o} \delta_{h} + { \bm \beta^\top \bm \delta} + \varepsilon_{y,d,v},
\end{align}
where $D_{y,d,v}$ is the  travel demand of vehicle type $v$  during the toll surcharge period (i.e., between 6 AM and 10 PM) on day $d$ in year $y$, $\alpha_y$ is the fixed effect of year $y$, $\delta_{v}$, $\delta_{h}$, and $\delta_{o}$ are dummy variables for vehicle type (1 if the vehicle type is a passenger vehicle, 0 otherwise), weekend/holiday (1 if day $d$ is a weekend or holiday), and Olympic/Paralympic period (1 if day $d$ is during the Olympic and Paralympic period, 0 otherwise), respectively. $\beta_i$ is the parameter of variable $i$ to be estimated, $\bm \beta$ and $ \bm \delta$ are vectors of other parameters and variables, such as day-of-week dummy variables,  and $\varepsilon_{y,d,v}$ is the error term of vehicle type $v$ on day $d$ in year $y$. By taking the logarithm for $D_{y,d,v}$ and $ P_{y, d, v}$, the parameters $\beta_{p}$ and $\beta_{p} + \beta_{ph}$ can be interpreted as the passenger vehicles' toll price elasticity for weekdays and weekends/holidays, respectively. Because the changes in traffic demand owing to toll surcharges during the Olympic and Paralympic Games can be captured by $\beta_{p}$ and $\beta_{p} + \beta_{ph}$,  the parameters  $\beta_{o}$ and $\beta_{o} + \beta_{oh}$ can be seen as the Olympic and Paralympic effect without tolling for weekdays and weekends/holidays. Therefore,  $\beta_{p}$ and $\beta_{p} + \beta_{ph}$ characterize the toll effect, while $\beta_{o}$ and $\beta_{o} + \beta_{oh}$ represent the Olympics effect for weekdays and weekends/holidays, respectively.

\subsection{Zone-based model: zone-based demand function}
We then investigate  spatial heterogeneity  by formulating the  demand function for each zone in the same manner as Eq. (\ref{eq:demand_function}).

\begin{align}\label{eq:demand_function_spatial}
\ln D^{z}_{y,d,v} = \alpha^{z}_y + \beta^{z}_y \delta_{v} + \beta^{z}_{p} \delta_{v} \ln P^{z}_{y, d, v} + \beta^{z}_{ph}\delta_{v} \delta_{h} \ln P^{z}_{y, d, v} + \beta^{z}_{o} \delta_{o} +  \beta^{z}_{oh} \delta_{o} \delta_{h} +  {\bm \beta^{z\top} \bm \delta^{z}} + \varepsilon^{z}_{y,d,v}.
\end{align}
The superscript ``$z$" represents the origin zone where traffic demand is generated. Thus, $D^{z}_{y,d,v}$ is travel demand generated during the toll surcharge period in zone $z$ for vehicle type $v$ on day $d$ in year $y$. We utilize the same variables as in  Eq. (\ref{eq:demand_function}).

\subsection{First-difference estimator}
\begin{figure}[!t]
\begin{minipage} {0.5\columnwidth}
  \centering
  \includegraphics[width=\textwidth]{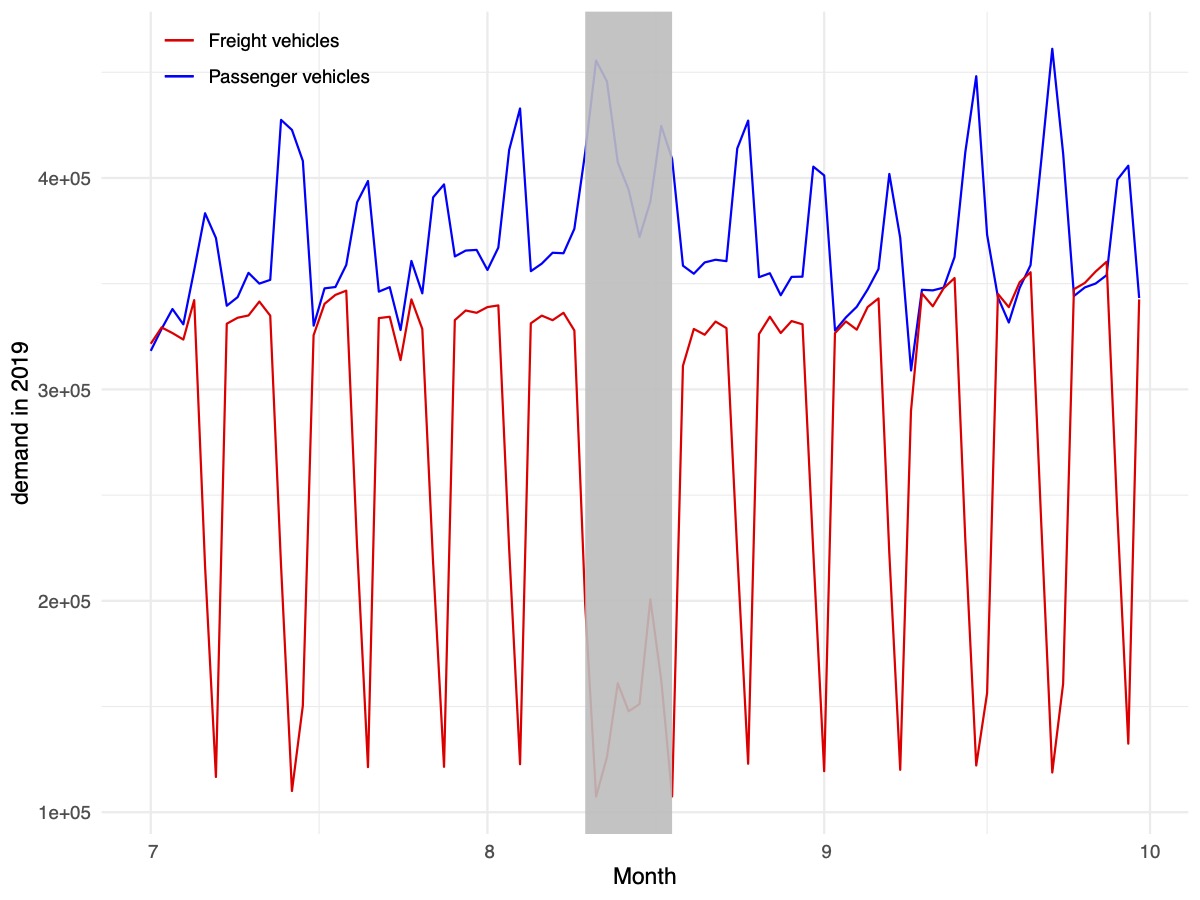}
  \caption{Daily demand of passenger and freight vehicles in 2019. The gray-colored area indicates the excluded dates. }\label{fig:demand}
\end{minipage}
\begin{minipage} {0.5\columnwidth}
  \centering
  \includegraphics[width=\textwidth]{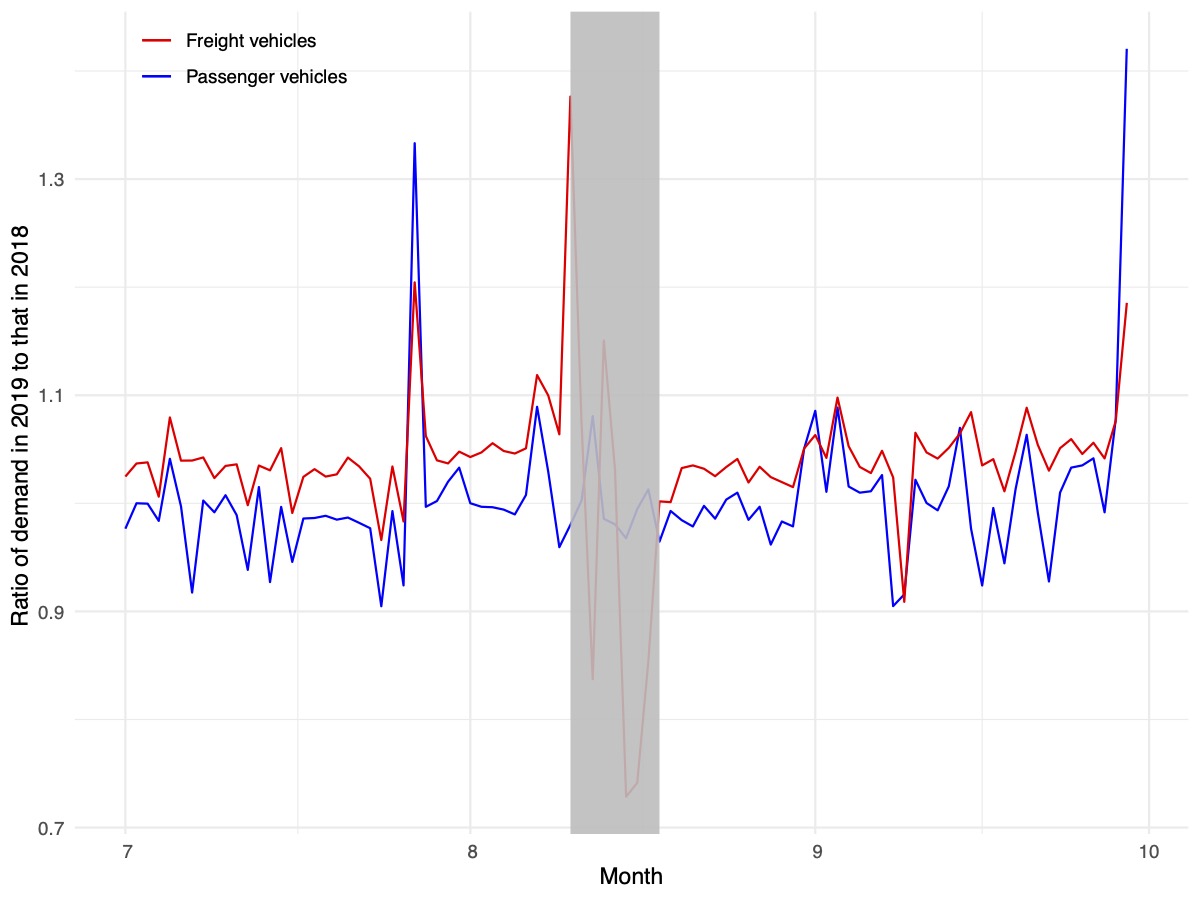}
  \caption{The ratio of daily demand for passenger and freight vehicles in 2019 to that in 2018. The gray-colored area indicates the excluded dates. }\label{fig:demand_ratio}
\end{minipage}
\end{figure}

As shown in Fig. \ref{fig:demand}, the raw  demand of passenger and freight vehicles does not satisfy the parallel trend assumption. Instead, we utilize a first-difference estimator. By taking the difference between Eq. (\ref{eq:demand_function}) for year $y$ and year $y-1$, we obtain 

\begin{align}\label{eq:FDmodel}
\ln \frac{D_{y,d,v}}{D_{y-1,d,v}}  = \alpha + \Delta\alpha_{2021} \delta_{{2021}} + \Delta \beta_{2021, v}  \delta_{{2021}} \delta_{v} + \beta_{p} \delta_{v} \ln \frac{P_{y, d, v}}{P_{y-1, d, v}} + \beta_{ph} \delta_{v} \delta_{h} \ln  \frac{P_{y, d, v}}{P_{y-1, d, v}} + \beta_{o} \delta_{o} +  \beta_{oh} \delta_{o} \delta_{h} +  \Delta \varepsilon_{y,d,v},
\end{align}
where $\Delta\alpha_{2021}$ is the fixed effect of the year 2021,  $\Delta \beta_{2021, v} $ is the fixed effect of passenger vehicles in the year 2021, and $\Delta \varepsilon_{y,d,i}$ is the error term. The LHS of Eq. (\ref{eq:FDmodel}) represents the logarithm of the ratio of demand for vehicle type $v$ on day $d$ in year $y$ to the demand  of vehicle type $v$ on day $d$ in year $y-1$. 

\begin{figure}[t!]
    \begin{minipage}{0.5\columnwidth}
      \centering
      \includegraphics[width=\textwidth]{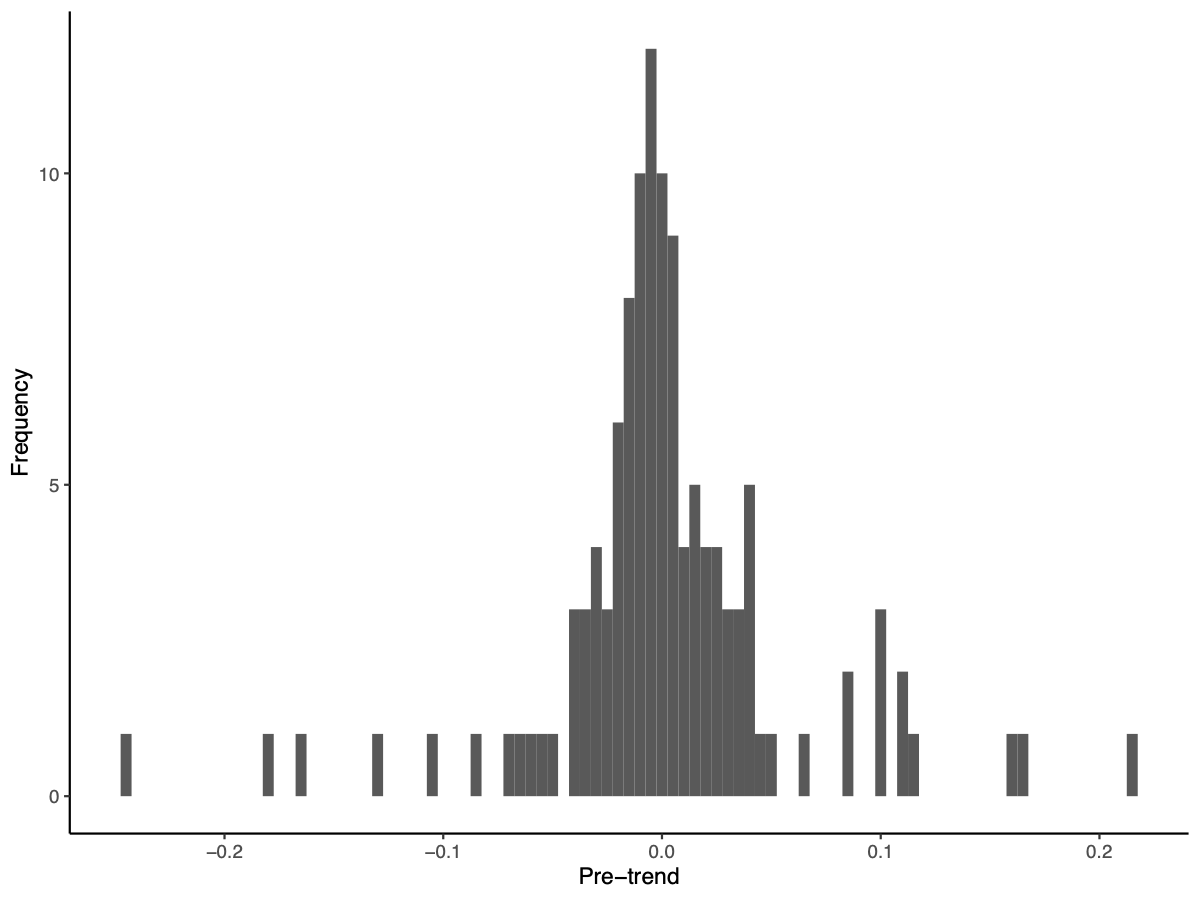}
      \caption{Histogram of the estimated pre-trend }\label{fig:trend} 
    \end{minipage}
    \begin{minipage}{0.5\columnwidth}
      \centering
      \includegraphics[width=\textwidth]{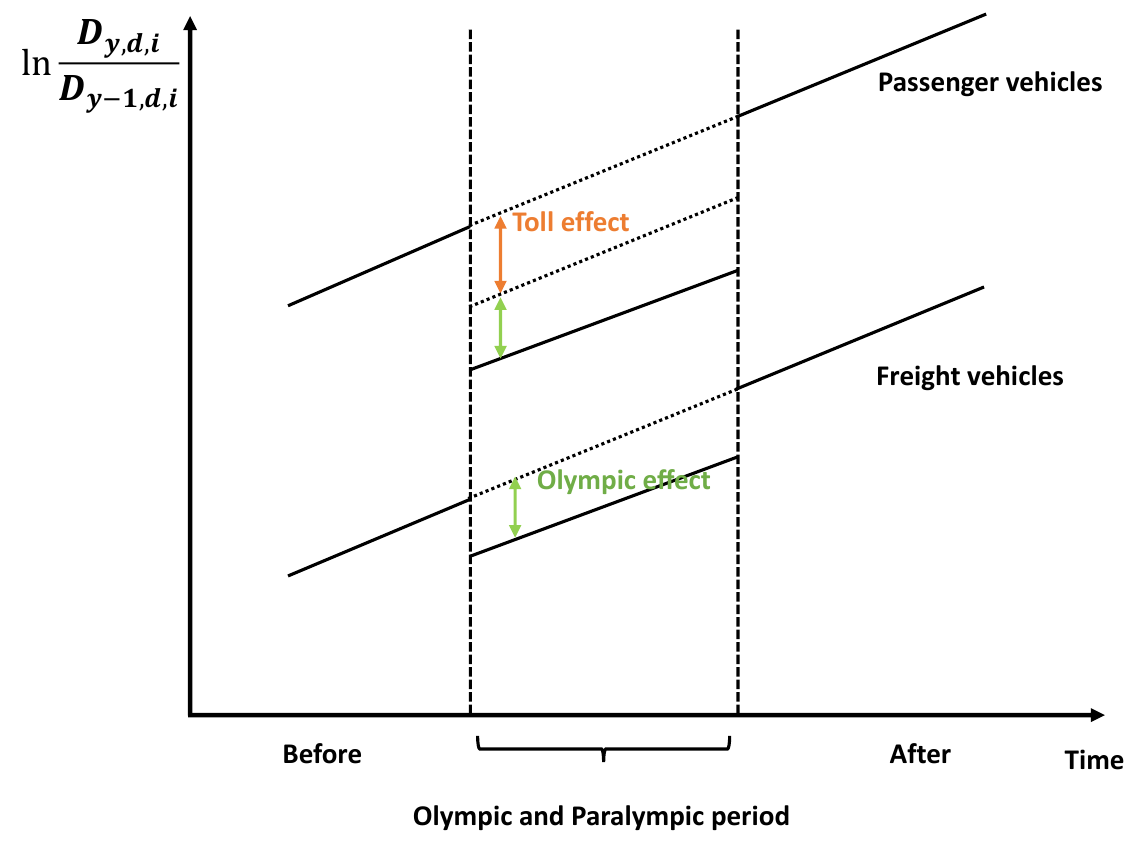}
      \caption{The graphical concept of the DID method in this study}\label{fig:DID}        
    \end{minipage}
\end{figure}

As depicted in Fig. \ref{fig:demand_ratio}, the trend  of the passenger and freight vehicles' ratio appears parallel.  The estimated pre-trend  result \footnote{The pre-trend $\Delta T_{y,d}$ at day $d$ in year $y$ can be estimated by $\Delta T_{y,d} = [Y_{y,d,v} - Y_{y,d+1,v} | (\delta_v, \delta_o)=(1,0)] - [Y_{y,d,v} - Y_{y,d+1,v} | (\delta_v, \delta_o)=(0,0)]$ where $Y_{y,d,v} = \ln\frac{D_{y,d,v}}{D_{y-1,d,v}} $} in Fig. \ref{fig:trend} indicates that most samples are centered around zero, suggesting that the parallel trend assumption holds. Not only does the parallel trend assumption hold, but the first-difference estimator can also address omitted variable bias. By applying Pooled-OLS regression, we estimate the parameters.

The concept of our DID method is illustrated in Fig. \ref{fig:DID}. Before and after the Olympic Games, the trend remains parallel. Because toll surcharges were not applied to freight vehicles, only the Olympics effect  ($\beta_{o}$ for weekdays and $\beta_{o} +\beta_{oh}  $ for weekends/holidays) affects the demand ratio of freight vehicles. Conversely, the demand ratio of passenger vehicles is also affected by the toll effect ($\beta_{p} $ for weekdays and $\beta_{p} 
 + \beta_{ph} $ for weekends/holidays) in addition to the Olympics effect.

 There are 13 date pairs where the date is a holiday in year $y$ but not in year $y-1$, and vice versa. There are also two pairs of dates in cases where one of the dates coincides with  a significant typhoon event (September 9th, 2019). Thus, we exclude these samples to maintain the parallel trend assumption. 

{Although we demonstrate that the parallel trend assumption holds, the same  Olympics effect on passenger and freight vehicle demand is not immediately  apparent. To investigate a similar effect to the Olympic effect, we leverage the data from the Olympic trials on July 24th and 26th 2019. On these dates, traffic control measures such as entry bans, and teleworking trials for the Tokyo 2020 Olympic Games were implemented, though the toll surcharges were not in place. We can reasonably expect a similar effect to the Olympic effect during these trials. When estimating the pre-trends for these dates, we find them to be 0.020 and 0.021, respectively. This suggests that the impact of the trials on passenger and freight vehicle demand can be considered equivalent.  Consequently, we can assume  an equal impact of the Olympics effect on passenger and freight vehicle demand. }

{Note that there are several ways to design a quasi-experiment for DID methods during the Olympic games. One possible approach is to separate passenger car trip data into those occurring during the toll surcharge period (6 AM to 10 PM) as the treatment group and those occurring outside the toll surcharge period (10 PM to 6 AM) as the control group. However, the same Olympic effect is not evident in this case, and we do not have trip data, such as that collected during the trials, to verify the presence of the same effect.  Since the data from the Olympic trials demonstrated that both passenger and freight vehicles were subject to the equivalent Olympic effect, we assume this assumption holds.}

\section{Results}

\subsection{General model}
Table \ref{table:numerical_result_whole} shows the regression results of the general model. An adjusted R-squared of 0.721 indicates high model accuracy, and all variables are statistically significant.

We observe that the effect of the year 2021  for passenger and freight vehicles is -0.0635 and -0.1182, respectively. The negative effect of the year 2021 indicates a decrease in the total demand for the Metropolitan expressway compared to 2019, which could be mainly due to the COVID-19 pandemic \citep{dantsuji2023understanding}. Passenger and freight vehicle demand decreased by  $6.1 \%$ and $11.1 \%$, respectively.

The magnitudes of the passenger vehicles' toll elasticity for weekdays and weekends/holidays are 0.345 and 0.615, respectively. The toll elasticity for weekdays aligns with a previous study on the Metropolitan expressway \citep{fukuda_shiroma}, which reported a range from 0.2 to 0.3. Moreover,  the higher elasticity for weekends/holidays than weekdays is consistent with previous studies \cite[e.g.,][]{holguin2005evaluation}. This could be because  weekday trips are primarily for business purposes, while those on weekends/holidays are more likely for leisure. These elasticities suggest that the toll surcharges led to a reduction of $25.0$ \% for weekdays and $36.8$ \% for weekends/holidays in passenger vehicle demand on the Metropolitan expressway\footnote{These changes in demand are estimated by $\left(\left(\frac{\bar{P}_{2021}}{\bar{P}_{2019}}\right)^{\beta_p + \beta_{ph} \delta_h} - 1 \right) \times 100$ where $\bar{P}_{2021}$ is the average toll price of passenger vehicles during the Olympic Games period in 2021 (1803.8 JPY and 1844.3 JPY for weekdays and weekends/holidays, respectively) and $\bar{P}_{2019}$ is the average toll price of passenger vehicles during the same period as the Olympic Games in 2019 (784.4JPY and 874.1 JPY for weekdays and weekends/holidays, respectively).}.

The estimated Olympic effect for weekdays is $0.0447$, suggesting a notable $4.6 \%$ surge in demand on weekdays during the Olympic Games, excluding the toll effect.  Conversely, we observe a contrasting Olympics effect for weekends/holidays (-0.0292), which suggests a $2.9 \%$ decrease on weekends/holidays. Although TDM strategies promote changes in travel behavior (TDM effect), such as altering destinations and reducing travel during the Games, other effects increase demand (induced effect), such as the demand  induced by congestion mitigation through TDM strategies. The Olympics effect encompasses all these effects. The outcome suggests that the induced effect surpasses the TDM effect on weekdays, although the TDM effect remains significant on weekends/holidays. 

\begin{table}[t!]
  \caption{Regression result for the general model}
  \label{table:numerical_result_whole}
  \centering
  \begin{tabular*}{15cm}{@{\extracolsep{\fill}}lcccc}
    Variable & Coefficient & SE & t value & p value \\  
    \hline
    Intercept & 0.0229 & 0.006 & 3.979 & 0.000 \\ 
    Year 2021 effect & -0.1182 & 0.012 & -9.587 &  0.000 \\
    Year 2021 effect * vehicle type & 0.0547 & 0.016 & 3.456 &  0.000 \\
    Toll effect & -0.3452 & 0.034 & -10.285 &  0.000 \\
    Toll effect * weekend/holiday & -0.2699 & 0.051 & -5.332 & 0.000 \\
    Olympics effect & 0.0447 & 0.019 & 2.310 & 0.021\\
    Olympics effect * weekend/holiday & -0.0739 & 0.028 &-2.672 &  0.008 \\
    \hline 
    &&& $R^2$ & 0.726 \\
    &&& Adjusted $R^2$ & 0.721 \\
    &&& Sample size & 330 
  \end{tabular*} 
\end{table}

\subsection{Zone-based model}

Next, we explore spatial heterogeneity. The results of the zone-based model are shown in Table \ref{table:numerical_result_spatial_hetero}.  R-squared values range from 0.463 to 0.814, demonstrating a high level of model accuracy. {Note that given the diversity in zone-specific characteristics, it is understandable that not all zones would yield highly accurate models in terms of R-squared. As the primary aim of the zone-based model is to explore the variability across different zones, rather than achieving highly precise predictions for each zone, the models do not necessarily need to be perfectly accurate for all zones to provide valuable insights into spatial heterogeneity.}

\begin{figure}[t!]
\begin{minipage} {0.5\columnwidth}
  \centering
  \includegraphics[width=\textwidth]{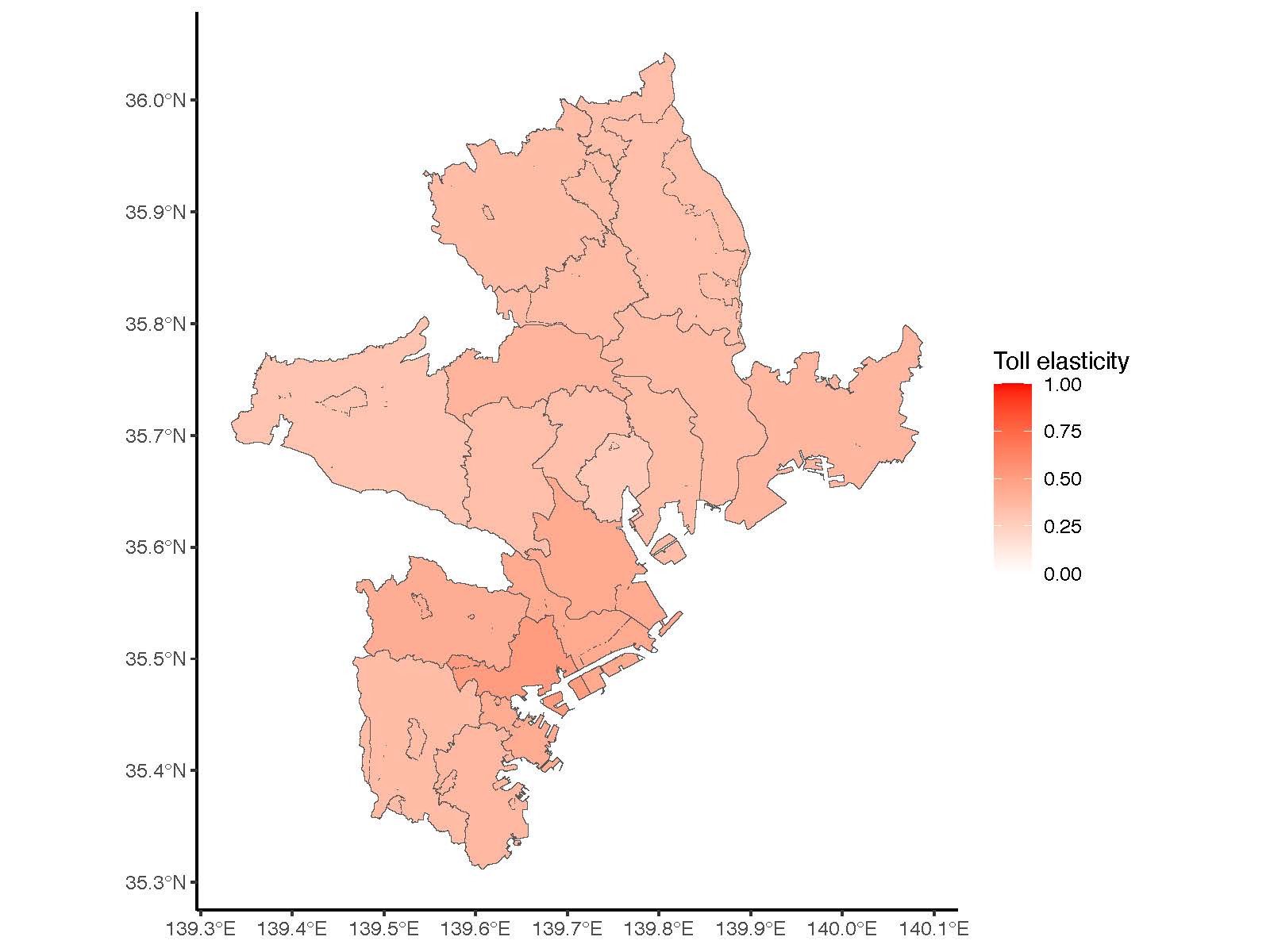}
  \caption{Spatial heterogeneity in toll elasticity (weekdays)}\label{fig:toll_elasticity_weekday}
\end{minipage}
\begin{minipage} {0.5\columnwidth}
  \centering
  \includegraphics[width=\textwidth]{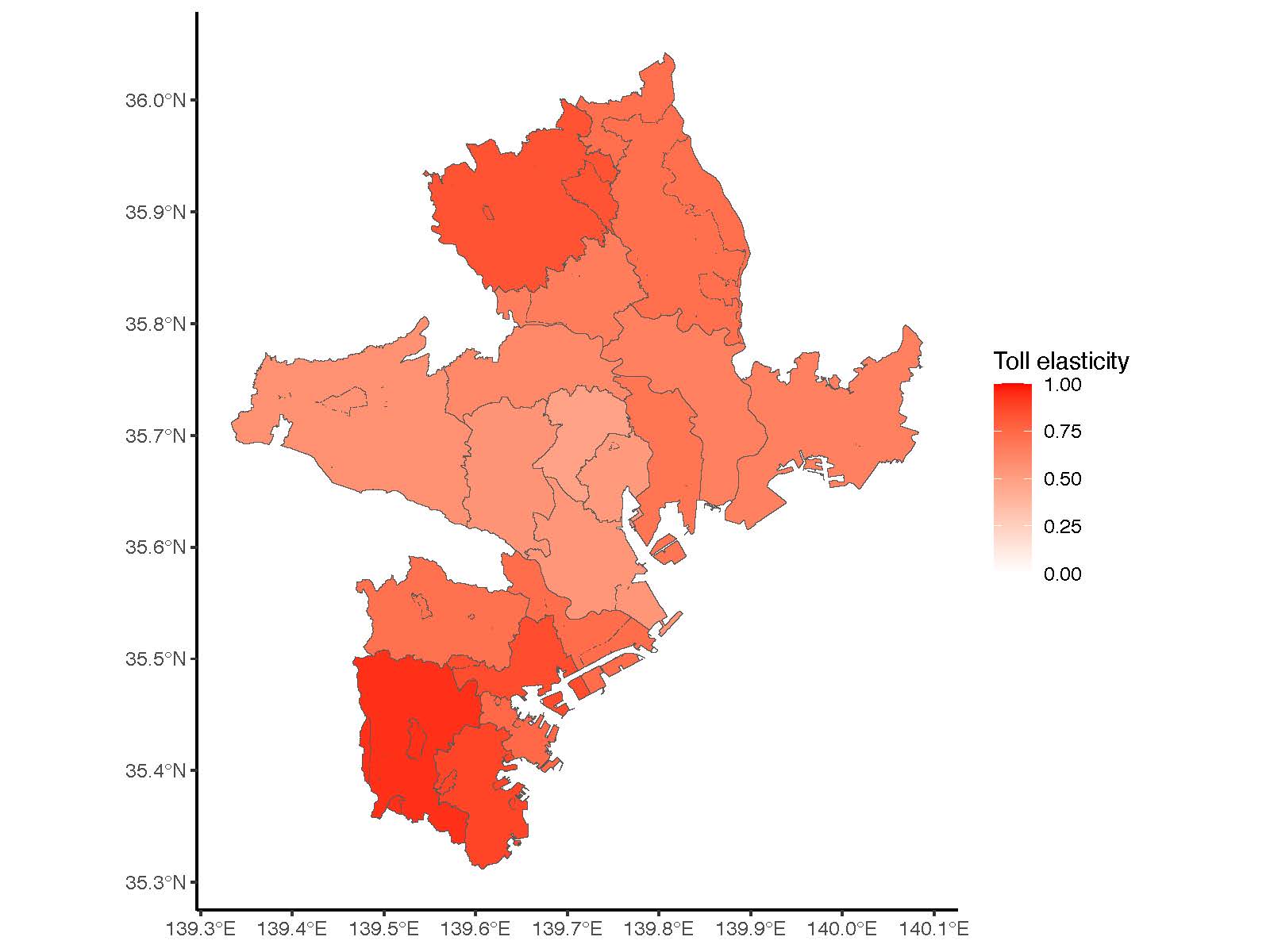}
  \caption{Spatial heterogeneity in toll elasticity (weekends/holidays)}\label{fig:toll_elasticity_holiday}
\end{minipage}
\end{figure}

All variables pertaining to the toll effect and the interaction term of the toll effect with weekends/holidays are significant. The toll elasticity ranges from 0.283 to 0.509 for weekdays and from 0.484 to 0.935 for weekends/holidays. Figures \ref{fig:toll_elasticity_weekday} and \ref{fig:toll_elasticity_holiday} show the spatial heterogeneity in toll elasticity for weekdays and weekends/holidays, respectively.  The higher toll elasticity observed for weekends/holidays aligns with the results of the general model. The narrower range of toll elasticity for weekdays may stem from differences in trip purposes because drivers with business purposes are typically less sensitive to price changes than those with private purposes.  {Notably, the toll elasticity for central zones tends to be lower than that for outer zones, especially during weekends/holidays. } This trend can be attributed to wealthier individuals, who are less price sensitive, being more likely to reside in central zones  (Fig 2 in \cite{tabuchi2019rich}), such as zones 1 -- 4. 

\begin{table}[p]
  \caption{Regression results for zone-based model}
  \label{table:numerical_result_spatial_hetero}
  \centering
      \renewcommand{\arraystretch}{1} 	
    \resizebox{\textwidth}{!}{%
  \begin{tabular}{llllllllllllllllll}
   Zone & Intercept & Year 2021 & \shortstack{Year 2021 * \\  vtype }  & Toll & \shortstack{Toll * \\ weekend/holiday } & Olympics & \shortstack{Olympics * \\ weekend/holiday }  & R$^2$  \\
  \hline  \hline 
  \multirow{2}{*}{1}	&	0.022***	&	-0.253***	&	0.200***	&	-0.283***	&	-0.238***	&	0.002	&	-0.011	&	\multirow{2}{*}{0.805}	\\
    &	(4.187)	&	(-22.355)	&	(13.753)	&	(-10.011)	&	(-5.631)	&	(0.137)	&	(-0.448)	&		\\\hline
  \multirow{2}{*}{2}	&	0.043***	&	-0.169***	&	0.042**	&	-0.438***	&	-0.100*	&	0.088***	&	-0.123***	&	\multirow{2}{*}{0.763}	\\
    &	(7.129)	&	(-12.789)	&	(2.459)	&	(-12.359)	&	(-1.912)	&	(4.308)	&	(-4.214)	&		\\\hline
  \multirow{2}{*}{3}	&	0.040***	&	-0.195***	&	0.163***	&	-0.331***	&	-0.153***	&	0.118***	&	-0.121***	&	\multirow{2}{*}{0.708}	\\
    &	(7.387)	&	(-16.678)	&	(11.050)	&	(-11.863)	&	(-3.652)	&	(6.442)	&	(-4.645)	&		\\\hline
  \multirow{2}{*}{4}	&	0.058***	&	-0.238***	&	0.074***	&	-0.344***	&	-0.351***	&	0.058***	&	0.027	&	\multirow{2}{*}{0.814}	\\
    &	(9.784)	&	(-18.776)	&	(4.492)	&	(-10.651)	&	(-7.383)	&	(2.897)	&	(0.935)	&		\\\hline
  \multirow{2}{*}{5}	&	0.006	&	-0.111***	&	0.061***	&	-0.341***	&	-0.206***	&	-0.046**	&	0.029	&	\multirow{2}{*}{0.724}	\\
    &	(0.927)	&	(-8.259)	&	(3.518)	&	(-10.140)	&	(-4.030)	&	(-2.164)	&	(0.941)	&		\\\hline
  \multirow{2}{*}{6}	&	0.051***	&	-0.077***	&	0.034**	&	-0.389***	&	-0.218***	&	-0.031	&	-0.055*	&	\multirow{2}{*}{0.768}	\\
    &	(8.114)	&	(-5.785)	&	(2.001)	&	(-11.378)	&	(-4.321)	&	(-1.480)	&	(-1.838)	&		\\\hline
  \multirow{2}{*}{7}	&	0.038***	&	-0.011	&	0.007	&	-0.355***	&	-0.290***	&	0.039*	&	-0.029	&	\multirow{2}{*}{0.672}	\\
    &	(6.264)	&	(-0.799)	&	(0.412)	&	(-10.493)	&	(-5.782)	&	(1.871)	&	(-0.981)	&		\\\hline
  \multirow{2}{*}{8}	&	-0.016**	&	-0.033*	&	0.017	&	-0.310***	&	-0.253***	&	0.124***	&	-0.162***	&	\multirow{2}{*}{0.463}	\\
    &	(-1.999)	&	(-1.945)	&	(0.797)	&	(-6.677)	&	(-3.619)	&	(4.611)	&	(-4.218)	&		\\\hline
  \multirow{2}{*}{9}	&	0.067***	&	-0.207***	&	0.072***	&	-0.432***	&	-0.315***	&	0.289***	&	-0.273***	&	\multirow{2}{*}{0.531}	\\
    &	(7.162)	&	(-10.273)	&	(2.737)	&	(-5.938)	&	(-3.034)	&	(9.136)	&	(-6.043)	&		\\\hline
  \multirow{2}{*}{10}	&	0.042***	&	-0.070***	&	0.035	&	-0.371***	&	-0.506***	&	-0.208***	&	0.029	&	\multirow{2}{*}{0.650}	\\
    &	(4.173)	&	(-3.256)	&	(1.260)	&	(-4.534)	&	(-4.411)	&	(-6.202)	&	(0.604)	&		\\\hline
  \multirow{2}{*}{11}	&	-0.004	&	-0.103***	&	-0.003	&	-0.351***	&	-0.584***	&	0.088***	&	-0.133***	&	\multirow{2}{*}{0.621}	\\
    &	(-0.487)	&	(-5.631)	&	(-0.118)	&	(-5.365)	&	(-6.307)	&	(3.079)	&	(-3.243)	&		\\\hline
  \multirow{2}{*}{12}	&	0.078***	&	0.551***	&	-0.175***	&	-0.424***	&	-0.287***	&	0.102***	&	-0.092**	&	\multirow{2}{*}{0.795}	\\
    &	(7.964)	&	(26.246)	&	(-5.876)	&	(-5.609)	&	(-2.895)	&	(3.106)	&	(-1.979)	&		\\\hline
  \multirow{2}{*}{13}	&	0.008	&	-0.069***	&	0.104***	&	-0.509***	&	-0.346***	&	0.052**	&	-0.140***	&	\multirow{2}{*}{0.642}	\\
    &	(1.143)	&	(-4.478)	&	(5.155)	&	(-9.664)	&	(-4.633)	&	(2.156)	&	(-4.091)	&		\\\hline
  \multirow{2}{*}{14}	&	0.040***	&	-0.084***	&	0.119***	&	-0.441***	&	-0.288***	&	-0.059*	&	-0.038	&	\multirow{2}{*}{0.562}	\\
    &	(4.177)	&	(-4.163)	&	(4.451)	&	(-7.259)	&	(-3.293)	&	(-1.853)	&	(-0.843)	&		\\\hline
  \multirow{2}{*}{15}	&	-0.017**	&	0.009	&	-0.072***	&	-0.355***	&	-0.300***	&	0.097***	&	-0.120***	&	\multirow{2}{*}{0.587}	\\
    &	(-2.284)	&	(0.540)	&	(-3.557)	&	(-7.182)	&	(-4.123)	&	(3.879)	&	(-3.364)	&		\\\hline
  \multirow{2}{*}{16}	&	-0.013*	&	-0.001	&	-0.030	&	-0.339***	&	-0.380***	&	0.036	&	-0.073**	&	\multirow{2}{*}{0.615}	\\
    &	(-1.804)	&	(-0.063)	&	(-1.480)	&	(-7.308)	&	(-5.544)	&	(1.468)	&	(-2.088)	&		\\\hline
  \multirow{2}{*}{17}	&	-0.003	&	-0.170***	&	-0.005	&	-0.377***	&	-0.267***	&	0.100***	&	-0.203***	&	\multirow{2}{*}{0.668}	\\
    &	(-0.381)	&	(-9.358)	&	(-0.227)	&	(-6.465)	&	(-3.143)	&	(3.499)	&	(-4.972)	&		\\\hline
  \multirow{2}{*}{18}	&	-0.026***	&	0.042**	&	0.007	&	-0.349***	&	-0.481***	&	0.134***	&	-0.203***	&	\multirow{2}{*}{0.502}	\\
    &	(-2.992)	&	(2.275)	&	(0.300)	&	(-5.474)	&	(-5.228)	&	(4.636)	&	(-4.920)	&		\\\hline \hline

  \hline 
  \end{tabular} }
    \raggedleft
   {\it  Note}:  t value in parentheses, and *$p<0.1$, **$p<0.05$, ***$p<0.01$
\end{table}

The variables related to the Olympics effect and its interaction with weekends/holidays are found to be insignificant and/or negative in certain zones. This result may be attributed to the real-time implementation of TDM strategies such as entry and lane closures. For instance, when an entry is closed within a zone, drivers may divert their entry to an alternative entry in another zone. Consequently, this dynamics leads to positive and negative effects in zones where entries are closed and  alternative entries are located. Owing to these intricate behavioral changes, the resulting combination of positive/negative and significant/insignificant Olympics effects, complicates the discussion of spatial heterogeneity in the Olympics effect. Nevertheless, we observe a positive Olympics effect at the aggregate level of the entire Metropolitan expressway, as indicated by the results of the general model.

\section{Conclusions and recommendations}
This paper evaluated the effects of toll surcharges and other Olympics-related factors on traffic demand on the Metropolitan expressway during the Tokyo 2020 Olympic and Paralympic Games. Utilizing panel data derived from longitudinal trip records on the expressway, we constructed a demand function and employed the DID method to estimate respective effects, leveraging the absence of toll surcharges for freight vehicles. Our findings indicate that toll surcharges resulted in a reduction of $25.0 \%$ for weekdays and $36.8 \%$ for weekends/holidays in traffic demand on the Metropolitan expressway. The estimated toll elasticities are $0.345$ for weekdays and $0.615$ for weekends/holidays.

We examined spatial heterogeneity in toll elasticity using the zone-based demand function. While toll elasticities are relatively homogeneous in space for weekdays ($0.283$ to $0.509$), significant heterogeneity was observed for weekends/holidays ($0.484$ to $0.935$). Higher toll elasticities for weekends/holidays may be attributed to differing trip purposes (e.g, business on weekdays and leisure on weekends/holidays), while the pronounced heterogeneity for weekends/holidays could stem from income disparities among zones.  Our results show that toll surcharges reduce passenger vehicle traffic demand on weekdays and weekends/holidays.

We also found that  traffic demand increased by $4.6 \%$ on weekdays  and decreased by $2.9 \%$ on weekends/holidays in the absence of toll surcharges. Despite implementing TDM strategies, such as promotional campaigns initiated before the Olympic Games, our results suggest that induced demand exceeded the reduction in demand from other TDM strategies on weekdays. {While existing literature frequently discusses the positive effects of travel demand management during the Olympic Games, the inverse impacts are rarely explored, despite their importance for effective TDM implementation \citep{lee1999induced}. We contributed by highlighting this aspect. }

{There are several important lessons learned from the Tokyo 2020 Olympic and Paralympic Games that can inform future transportation policies, especially during mega events. We demonstrated that the tolling strategy implemented during the Games was effective in reducing traffic demand. While existing studies have already showed the effectiveness of daily pricing strategies, such as congestion pricing, in managing traffic flow under normal conditions \cite[e.g.,][]{borjesson2012stockholm}, our findings show that toll surcharges can also be highly effective in managing short-term traffic surges during mega events. This is particularly relevant for host cities of international events where temporary but significant increases in traffic are expected. }

{Moreover, We revealed the spatial and temporal (weekdays vs weekends) heterogeity in the impact of toll surcharges. This suggests that a uniform toll surcharge may not be optimal. Future tolling schemes could be more effective if they incorporate zone-based pricing \cite[e.g., ][]{zheng2020area} and time-dependent pricing \cite[e.g.,][]{dantsuji2021simulation}. For the successful implementation of these advanced tolling schemes, understanding toll elasticity is crucial. Therefore, the estimated toll elasticities are valuable for such future endeavors. }

{In addition to tolling strategies, it is important to consider the broader context of TDM during mega events. \cite{currie2014travel} have highlighted the ``big scare" effect, where TDM program warnings, media attention, and unplanned events before the Games influenced travel behaviors. However, our findings indicate that induced and/or unchanged demands were more significant on weekdays. Because the traffic conditions on the Metropolitan expressway were less congested during the Games, it is plausible that more drivers preferred using it over highways to save travel time. This shift in behavior highlights the complexity of predicting and managing traffic flows during mega events.}

{Given the complex and sometimes unpredictable nature of these effects, pre-determined strategies might have limited or even unintended consequences on transportation systems. This underscores the need for more flexible and adaptive approaches to traffic management. As real-time data becomes increasingly accessible, online learning-based control strategies that leverage digital twin technology \cite[e.g.,][]{dantsuji2024hybrid} offer a promising alternative for managing transportation systems. These approaches can continuously monitor traffic conditions, learn from real-time data, and adjust control strategies dynamically, providing a more responsive and effective means of handling traffic during mega events. }

{The experience of the Tokyo 2020 Games provides valuable insights into the potential and limitations of traditional traffic management strategies. Moving forward, integrating adaptive, data-driven approaches will be essential for managing transportation systems during both daily operations and mega events.}

The current research can be extended in a few directions. First, while we assessed the impacts of toll surcharges and other Olympics-related factors separately. Estimating the causal effects of additional TDM strategies, such as the promotional campaigns, would be beneficial. {Second, investigating the impact of TDM measures on different driver characteristics (e.g., Tokyo residents vs. residents from other areas) and travel modes, such as taxis, is an interesting direction for this paper. This could be achieved by combining the current data with additional sources, such as taxi GPS data \citep{yang2017modeling}. } Third, traffic dynamics were expected to be significantly different from  usual during the Tokyo Olympics Games, as indicated in the previous Games \citep{mingjun2008comparison}. Analyzing the TDM effects on traffic dynamics would be an interesting direction for future research.

\section*{Declaration of Competing Interest }
None.

\section*{Acknowledgements}
We thank the Tokyo Metropolitan government for providing the data. We also thank Yuki Takayama, Daisuke Fukuda, Makoto Chikaraishi,  Wataru Nakanishi, Hideki Oka, and Takara Sakai for their valuable comments. This work was supported by JST ACT-X, Japan, Grant \#JPMJAX21AE and JSPS KAKENHI, Japan, Grant \#23K13422.

\section*{Declaration of generative AI and AI-assisted technologies in the writing process}
During the preparation of this work the authors used ChatGPT in order to improve the written text. After using this tool/service, the authors reviewed and edited the content as needed and take full responsibility for the content of the publication.

\bibliographystyle{elsarticle-harv}\bibliography{ref}
\end{document}